\documentclass[10pt,twocolumn,twoside]{IEEEtran}

\IEEEoverridecommandlockouts                              

\usepackage{enumerate}
\usepackage{amsmath,color}
\usepackage{amssymb}
\usepackage{amsfonts}
\usepackage{graphicx}
\usepackage{epstopdf}
\usepackage{amsmath}
\usepackage{cancel}
\usepackage{mathrsfs}
\usepackage{mathdots}
\usepackage{euscript}
\usepackage{amscd}
\usepackage{cite}
\usepackage{placeins}
\usepackage{algorithm2e}
\usepackage{tikz}
\usetikzlibrary{snakes,arrows,shapes}

\graphicspath{{images/}}

\newtheorem{thm}{Theorem}[section]
\newtheorem{ass}{Assumption}[section]
\newtheorem{defn}{Definition}[section]
\newtheorem{lem}{Lemma}[section]
\newtheorem{cor}{Corollary}[section]
\newtheorem{prob}{Problem}[section]
\newtheorem{prop}{Proposition}[section]
\newtheorem{rem}{Remark}[section]

\definecolor{Royalblue}{cmyk}{1,0.30,0.2,0.2}

\newcommand{\proof}{\noindent {\it Proof. }}

\newcommand{\qed}{\hfill $\Box$ \vskip 2ex}

\newcommand{\btheo}{\begin{thm}}
\newcommand{\bdeff}{\begin{deff}}
\newcommand{\bass}{\begin{ass}}
\newcommand{\blem}{\begin{lem}}
\newcommand{\bcor}{\begin{cor}}
\newcommand{\bprob}{\begin{prob}}
\newcommand{\bprop}{\begin{prop}}
\newcommand{\brem}{\begin{rem}}

\newcommand{\etheo}{\end{thm}}
\newcommand{\edeff}{\end{deff}}
\newcommand{\eass}{\end{ass}}
\newcommand{\elem}{\end{lem}}
\newcommand{\ecor}{\end{cor}}
\newcommand{\eprob}{\end{prob}}
\newcommand{\eprop}{\end{prop}}
\newcommand{\erem}{\end{rem}}
 
\newcommand{\pinv}{^{\dagger}}

\newcommand{\bmat}{\left[ \begin{matrix}}
\newcommand{\emat}{\end{matrix} \right]}

\newcommand{\script}[1]{\EuScript{#1}}

\newcommand{\Rbb}{\mathbb R}

\newcommand{\tp}{^{\top}}
\newcommand{\mtp}{^{-\top}}
\newcommand{\inv}{^{-1}}

\newcommand{\beq}{\begin{equation}}
\newcommand{\eeq}{\end{equation}}
\newcommand{\bea}{\begin{eqnarray}}
\newcommand{\eea}{\end{eqnarray}}

\newcommand{\bsea}{\begin{subeqnarray}}
\newcommand{\esea}{\end{subeqnarray}}
\newcommand{\nn}{\nonumber}

\begin{document}
\title{On the state space and dynamics selection in linear stochastic models: a spectral factorization approach
}
\author{Augusto Ferrante and Giorgio Picci 
\thanks{A. Ferrante and G. Picci    are with the
Department of Information Engineering, University of Padova, via
Gradenigo 6/B, 35131 Padova, Italy; e-mail:  
{\tt augusto@dei.unipd.it}, {\tt picci@dei.unipd.com}} 
}
\date{\today}
\maketitle

\begin{abstract}
 Matrix spectral factorization is traditionally  described as  finding spectral factors having a fixed analytic pole configuration. The classification of spectral factors then  involves studying the solutions of a certain algebraic Riccati equation which parametrizes  their zero structure. The pole structure of the spectral factors can be also parametrized in terms of solutions of another Riccati equation. We study the  relation between the solution sets of these two Riccati equations and describe the construction of general spectral factors which  involve both zero- and  pole-flipping on an arbitrary reference spectral factor.
  \end{abstract}


 \section{Introduction}
An important and widely used class of models in control engineering and signal processing describes an $m$-dimensional observed random signal $\{y(t)\}$ as output of a linear system driven by white noise:
\beq
\label{star}
\left\{\begin{array}{l}
x(t+1)=Ax(t)+Bw(t)\\
y(t)=Cx(t) +D w(t)
\end{array}
\right.
\eeq
where $A\in\Rbb^{n\times n}$, $B\in\Rbb^{n\times m}$, $C\in\Rbb^{m\times n}$, $D\in\Rbb^{m\times m}$,
$w$ is a normalized white noise.  The $n$-dimensional signal $x$ is the state vector.
The basic steps for the constructions of models of the form \eqref{star} from   observations of  $\{y(t)\}$   lead to the following three problems which in various forms permeate all  linear systems and control theory:
\begin{enumerate}
\item Estimate the spectral density $\Phi_y(z)$ of $y$,  see \cite{Ferrante-Masiero-Pavon-TAC-12,GEORGIOU_MAXIMUMENTROPY,GEORGIOU_RELATIVEENTROPY,BETA,OPTIMAL_PREDICTION_ZORZI_2014,ALPHA} and references therein.
\item
Compute a stochastically minimal\footnote{Stochastic minimality means that we are only interested in models of minimal complexity so that 
we only consider spectral factors $W(z)$ of minimal McMillan degree.} spectral factor of $\Phi_y(z)$, i.e. a matrix transfer function
$W(z)$ such that 
\beq\label{sfcator}
\Phi_y(z)=W(z)W\tp(z^{-1}),
\eeq
  see \cite{Baggio-Ferrante-TAC-16-2,Baggio-Ferrante-TAC-16} and references therein.
\item
 Fix a minimal realization  $W(z)=C(zI-A)^{-1}B+D$ to provide a parametrization of the model \eqref{star}.
 \end{enumerate}
The literature on these topics being enormous  we have chosen to quote only a few  recent papers in which one can find a more extensive bibliography.
The study  of   models \eqref{star} of the   signal $y$ without a priori constraints of causality or analiticity is exposed in the recent book \cite{LPBook}.  The objective of this paper is to continue  the analysis and study in more  depth the    relations among different  models  \eqref{star} which are in a sense {\em equivalent} as they serve to represent the same process but may have different system-theoretic  structure and properties.\\
Indeed,  representations  \eqref{star} have  several degrees of freedom. The most obvious (and least interesting) one 
is the choice of basis in the input and in the state space.
In particular, the matrices $A,B,C,D$ in step 3. are determined up to a transformation of the form
$T^{-1}AT,T^{-1}BU,CT,DU$ where $T$ is an arbitrary invertible matrix and $U$ is an arbitrary orthogonal matrix.
Once these degrees of freedom are factored out, we are left with   two   more interesting objects:\\
A. The state space as a coordinate free representative of a model \eqref{star}\\
B. The (dynamical)  causality structure (related in particular the choice of direction of the time arrow) of equivalent    models.\\
One of the key result of   stochastic realization theory  
(see \cite{LPBook}) is that these two choices correspond, respectively, to the {\em choice of zeros and poles}   of the spectral factor $W(z)$ in \eqref{sfcator}.
Each   pole configuration of the spectral factor corresponds to a certain causality structure  so that, once this configuration is fixed,  one is  left with the choice   of the zero structure of the spectral factor, which just means choosing  a (minimal) state space of the realization.

Matrix spectral factorization is traditionally  described as  finding spectral factors having a fixed analytic pole configuration  so that all corresponding models are causal, and classifying different models   corresponds to parametrizing all   possible zero structures of $W$. However, a zero structure fixes, independent of causality, a possible minimal state space\footnote{ 
We stress that the choice of the state space must not be confused with the choice of basis in $\Rbb^n$.} for $y$.
Hence, once  a minimal state space (i.e. the zero structure of $W$)  is fixed, there is a whole family of    possible  causality structures   which can be  parametrized by the allowed pole locations  of a spectral factor $W$.

If some minimal {\em reference} spectral factor is fixed, minimal spectral factorization can  be seen as a {\em zero- or pole- flipping} transformation performed on the reference factor.    In this paper we analyse the interplay between this two operations in relation to the solutions 
sets of two families of algebraic Riccati equations. We derive closed-form formulas that allow to compute the model corresponding to a given causality structure and state space.
This may be viewed as the completion of an endeavour first undertaken in \cite{Picci-P-94}  in continuous time but not pushed to the final consequences.  Here we shall address the discrete-time situation and give a complete solution.

Although  our main motivation is stochastic modelling, our contribution can also be viewed as 
related to   algebraic Riccati equations  and  to spectral factorization. Both have important applications in several areas of control, signal processing and system theory.

Some technical assumptions of this  paper  could probably be weakened    however most probably    at the expense of clarity. For pedagogical reasons we have decided to work in a setting which reduces complications to a minimum.

\section{ Background on spectral factorization and Algebraic Riccati equations}
Let $\Phi(z)$ be a $m\times m$ rational spectral density matrix of a regular stationary process, where regularity is meant in the sense explained in \cite[Sec. 6.8]{LPBook} and let
\beq\label{FSdiPartenza}
W(z):= C(zI-A)^{-1}B+D
\eeq
be a minimal realization of a minimal square spectral factor of $\Phi(z)$ so that $\Phi(z)= W(z)W(z)^{*}$, where $W(z)^{*}:= W(z^{-1})^{\top}$  is the conjugate transpose. By regularity  the   matrix $D$ is non singular,  \cite{Ferrante-P-P-02-LAA}; it will be assumed to be symmetric and positive definite: this rules out the   uninteresting degree of freedom corresponding to multiplying   a spectral factor on the right side  by a constant orthogonal matrix.

By regularity the {\em numerator matrix}  $\Gamma:=A-BD^{-1}C$ is non-singular (see Theorem 6.8.2 in \cite{LPBook}). In this paper we shall moreover assume that  both $A$ and $\Gamma$ are unmixed. Note in particular   that we do  not assume analiticity of $W(z)$ outside of the unit disk. For the relevant definitions and facts about spectral factorization in this context we shall refer to   Chap 16 of the book \cite{LPBook}. 
\begin{defn} {\em
Let $W_i(z)\,;\, i=1,2$ be minimal spectral factors of the same rational spectral density.
We shall say that  $W_1(z)$ and $W_2(z)$ have the {\em same pole  structure} if they admit 
a state space realization with the same  state transition matrix. Likewise, we say that $W_1(z)$ and $W_2(z)$ have the {\em same zero  structure} if they admit 
a state space realization with the same numerator matrix.}
\end{defn}
In classical spectral factorization one assumes that a the state matrix  $A$ has all eigenvalues inside the unit circle  and one aims at classifying all different minimal spectral factors having a fixed (analytic) pole structure, in terms of their zero structure, equivalently, in terms of invariant subspaces for the transpose of the numerator matrix $\Gamma$. It is well-known that this involves  the study of an algebraic Riccati equation. In the present context we have the following result, which has appeared  in several places in the literature. 
\begin{prop}\label{prpz}
{\em
Let $W_0(z):= C(zI-A)^{-1}B+D$ be a minimal realization of a square reference spectral factor. \begin{enumerate}
\item
There is a one-to-one correspondence between symmetric solutions of the homogeneous algebraic Riccati equation
\beq\label{ch-AREP}
P=\Gamma P\Gamma\tp -\Gamma PC\tp (DD\tp +CPC\tp)\inv CP\Gamma\tp
\eeq
 and  minimal spectral factors of $\Phi(z)$ having the same pole structure of $W_0(z)$. This correspondence is defined by the map assigning to each solution $P$ the spectral factor
\beq \label{WP}
W_{\small P}(z):= C(zI-A)^{-1}B_{\small P}+ D_{\small P}\,
\eeq
where  
\beq \label{PBD}
\begin{aligned}
& B_{\small P}:=(BD\tp +APC\tp ) (DD\tp +CPC\tp)^{-1/2}\,; \\
& D_{\small P}:= (DD\tp +CPC\tp)^{1/2}\,.
\end{aligned}
\eeq
 \item
 There is a one-to-one correspondence between symmetric solutions of
(\ref{ch-AREP}) and $\Gamma\tp$-invariant subspaces
which is  defined by the map assigning to each solution $P$ the $\Gamma\tp$-invariant subspace $\ker(P)$.
\end{enumerate}
}
\end{prop}
For a proof  we shall just refer the reader to  Corollary 16.5.7 and Lemma 16.5.8 in \cite{LPBook} where the equation differs by an inessential change of sign. A similar Riccati equation although in a different context  is studied in  \cite{Wimmer-06}.\footnote{Any solution $P$ can actually be seen as the difference say $X-X_0$ of two arbitrary solutions of an equivalent Riccati equation parametrizing the minimal spectral factors which is defined directly in terms of a  realization of $\Phi$ and does not  involve  a reference  spectral factor,   see \cite[Sect. 16.5]{LPBook}.  Here $X_0$ is kept fixed as a reference solution and $ \Gamma $ describes the zero structure of the reference spectral factor $W_0$.} \\

In particular, let $P_{+}$ be the unique non singular solution of \eqref{ch-AREP}, then  the corresponding $\Gamma^{\top}-$invariant subspace $\ker {P_{+}}$ is trivial and the zeros of $W_0(z)$  are all flipped to   reciprocal positions.  This  follows from standard Riccati theory. We shall denote the corresponding spectral factor by $W_+(z)$.\\
Zero-flipping  can also be visualized  as right-multiplication of $W_0(z)$ by a suitable square all-pass function so as to preserve minimality. The 	entailed factorization of $W_{\small P}(z)$ is in turn uniquely identified by the existence of a $\Gamma\tp-$invariant subspace \cite{Bart-G-K-84}. 

On the other hand, we have the following fact which describes the {\em pole-flipping} relation among spectral factors keeping a {\em fixed zero structure}. The result can be traced back to Theorem 16.4.2 of \cite{LPBook}.
\begin{prop}\label{prop} 
{\em
Let $W_0(z):= C(zI-A)^{-1}B+D$ be a minimal realization of a square reference spectral factor. 
\begin{enumerate}
\item
There is a one-to-one  correspondence between symmetric solutions of the algebraic Riccati equation
\beq\label{ch-AREQ}
Q = A\tp Q A - A\tp Q B\,(I+B\tp QB)^{-1} B\tp QA\,, 
\eeq
and  minimal normalized spectral factors  having the same zero structure of $W_0(z)$.
This correspondence is defined  by the map assigning to each solution $Q$ the spectral factor
\beq \label{WQ}
W_{\small Q}(z):= C_{\small Q} (zI-A_{\small Q})^{-1}B_Q+ D_Q,
\eeq
where
\beq \label{CDQ}
\begin{aligned}
&\Delta_{\small Q}:=\,I+B\tp QB\,,\\
&C_{\small Q}:= C-D\Delta_{\small Q}^{-1} B\tp QA\,,\\
&A_{\small Q}:=A -   B \Delta_{\small Q}^{-1} B\tp QA\,,\\
&B_Q:=B\Delta_{\small Q}^{-1/2}U\,,\\
&D_Q:=D\Delta_{\small Q}^{-1/2}U\,,
\end{aligned}
\eeq 

and $U$ is   the orthogonal matrix $$
U:=(D\Delta_{\small Q}^{-1/2})\tp ((D\Delta_{\small Q}^{-1/2})(D\Delta_{\small Q}^{-1/2})\tp)^{-1/2}
$$
which is selected in such a way that  $D_Q$ is symmetric and positive definite.
\item
There is a one to one correspondence between symmetric solutions of
(\ref{ch-AREQ}) and $A$-invariant subspaces
which is  defined by the map assigning to each solution $Q$ the $A$-invariant subspace $\ker(Q)$.
\end{enumerate}
}
\end{prop}

\proof
That the zero structures of $W_{\small Q}(z)$ and of $W_0(z)$ coincide is the content of Theorem 16.4.5 in \cite{LPBook}. The rest is readily checked.
\qed
\section{  Combining pole and zero flipping}

We want to understand the combination of zero- and pole-flipping leading to an arbitrary minimal square spectral factor  $W$. To this end let's consider the spectral factor $W_{\small Q}(z)$ defined in \eqref{WQ}
as a reference spectral factor and describe the zero-flipping process on $W_{\small Q}(z)$. 
  By direct computation we easily find that the numerator matrix of $W_{\small Q}(z)$ is 
the same of the numerator matrix of $W_0(z)$, i.e. the matrix $\Gamma$. Hence the  Riccati equation (\ref{ch-AREP})  corresponding to $W_{\small Q}(z)$   takes the form
\beq\label{ch-AREPq}
P_{\small Q}=\Gamma P_{\small Q} \Gamma\tp -\Gamma P_{\small Q}C_{\small Q}\tp (D_{\small Q}D_{\small Q}\tp +C_{\small Q}P_{\small Q}C_{\small Q}\tp)\inv C_{\small Q}P_{\small Q}\Gamma\tp
\eeq
where $C_{\small Q}$ is as  defined in \eqref{CDQ} and $D_{\small Q}:=D\Delta_{\small Q}^{-1/2}U$.
Notice  that, since equations (\ref{ch-AREP}) and (\ref{ch-AREPq}) involve the same matrix $\Gamma$ and each symmetric solution of either equation is uniquely attached to a $\Gamma\tp$-invariat subspace \cite{Willems-71},  the map assigning to each solution $P$ of  (\ref{ch-AREP}) the solution $P_{\small Q}$ of (\ref{ch-AREPq}) such that $\ker(P)=\ker(P_{\small Q})$ is a one to one correspondence between the set
${\mathcal P}$ and  the set  ${\mathcal P}_{\small Q}$ of symmetric solutions of \eqref{ch-AREP} and      (\ref{ch-AREPq}). 

  Our main contribution is to analyze the relations between ${\mathcal P}$ and  ${\mathcal P}_{\small Q}$ 
and to provide an explicit formula to compute the solution  ${\mathcal P}_{\small Q}$ from a give pair $P,Q$.
In this way once parametrized the solutions of (\ref{ch-AREP}) and \eqref{ch-AREQ}, we do not need to solve
(\ref{ch-AREPq}) and we have a closed-form formula for the spectral factor with assigned pole and zero structure, or equivalently for the model with assigned state-space and causality structure.  

We may of course consider a dual path to transform $W_0$ into $W$ by   taking instead the zero-flipped $W_{\small P}$ as a reference and flipping poles by considering a solution $Q_{\small P}$ of a Riccati equation similar to \eqref{ch-AREQ} so as to make the following diagram commutative:
\beq \label{CD}
\begin{CD}
 W_0        @> {P }     >>      W_{\small P}   \\
@V{Q}VV           @VV{Q_{\small P}}V  \\  
 W_{\small Q} @>>{P_{\small Q}}>  W
\end{CD}
\eeq
The resulting spectral factor should have been denoted $W_{PQ}$ but the simplified notation $W$ here should not be cause of confusion.

It  is well-known that  both (\ref{ch-AREP}) and \eqref{ch-AREPq} have  a unique non-singular solution   which we denote by $P_+$ and $P_{\small Q +}$, respectively. The relation between these two solutions is the content of the following lemma. 
\begin{lem}
{\em
The nonsingular solutions $P_{+}$ and $P_{\small Q +}$ are  related by the formula
\beq\label{Nonsing}
P_{\small Q +}^{-1}=Q + P_{+}^{-1}\,.
\eeq
}
\end{lem}
\proof
It is immediate to check that $P_{+}^{-1}$ is  the (unique) solution of the discrete-time Lyapunov equation
\beq
P_{+}^{-1}-\Gamma\tp P_{+}^{-1}\Gamma +C\tp D\mtp D\inv C=0.
\eeq
Similarly,
$P_{\small Q +}^{-1}$ is  the (unique) solution of the discrete-time Lyapunov equation
\beq
P_{\small Q +}^{-1}-\Gamma\tp P_{\small Q +}^{-1}\Gamma +C_{\small Q}\tp D_{\small Q}\mtp D_{\small Q}\inv C_{\small Q}=0.
\eeq
Therefore, the difference $\Delta:=P_{\small Q +}^{-1}-P_{+}^{-1}$ is the (unique) solution of the discrete-time Lyapunov equation
\beq\label{ch-eqforDelta}
\Delta -\Gamma\tp \Delta \Gamma +C_{\small Q}\tp D_{\small Q}\mtp D_{\small Q}\inv C_{\small Q}-C\tp D\mtp D\inv C=0.
\eeq
We now compute
\bea
\nn
R_{\small Q}&:=&C_{\small Q}\tp D_{\small Q}\mtp D_{\small Q}\inv C_{\small Q}-C\tp D\mtp D\inv C\\
\nn
&=&C\tp D\mtp\Delta_{\small Q} D\inv C +   A\tp Q B \Delta_{\small Q}^{-1} B\tp QA\\
\nn
&&
-C\tp D\mtp B\tp QA - A\tp Q BD\inv C \\
\nn
&&
-C\tp D\mtp D\inv C\\
\nn
&=&C\tp D\mtp B\tp QB D\inv C -C\tp D\mtp B\tp QA \\
\nn
&&
- A\tp Q BD\inv C +A\tp Q B \Delta_{\small Q}^{-1} B\tp QA\\
\nn
&=&C\tp D\mtp B\tp QB D\inv C -C\tp D\mtp B\tp QA \\
\nn
&&
- A\tp Q BD\inv C +A\tp Q  A-Q  \\
\nn
&=&\Gamma\tp Q  \Gamma -Q  
\eea
This equation, together with (\ref{ch-eqforDelta}) gives
\beq
\Delta -\Gamma\tp \Delta \Gamma =Q-\Gamma\tp Q \Gamma
\eeq
and, by uniqueness, 
$ \quad
\Delta:=P_{\small Q +}^{-1}-P_{+}^{-1}=Q,
\quad $
so that \eqref{Nonsing} follows.
\qed
 Since $\ker P_{+}= \{0\}$, all zeros of the corresponding spectral factor,   denoted by the symbol  $W_+(z)$,  are those of $W_0$ flipped to their reciprocals. The same happens for $W_{\small Q}$ whatever   solution $Q$ of \eqref{ch-AREQ} is  chosen. In particular, denoting  the nonsingular  solution of \eqref{ch-AREQ}  by $Q_+$, all poles of  the corresponding spectral factor, say $\bar{W}_0(z)$, will be the reciprocals of those of $W_0(z)$. The commutative diagram \eqref{CD} takes on the form
\[
\begin{CD}
 W_0        @> {P_+ }     >>      W_{+}   \\
@V{Q_+}VV           @VV{[Q_{\small P_+}]_+}V  \\ 
 \bar {W}_{0} @>>{[P_{\small  {Q_+} }]_+}>  \bar W_+
\end{CD}
\]
where $[Q_{\small P_+}]_+$ and $[P_{\small  {Q_+} }]_+$ are the invertible solutions of the Riccati equations which are respectively flipping the poles  of $W_{+} (z)$ and the zeros of $\bar W_{0} (z)$. Hence, both   poles and   zeros of $\bar W_+$ are the reciprocals of those of $W_0(z)$. This corresponds to   ``total'' flipping of singularities.

We would now like to derive an explicit formula generalizing  \eqref{Nonsing} to a generic solution $P_{\small Q}$ expressed as a function of $P$ and $Q$.  To this end we shall use the following lemma which is   a particular case of  \cite[Theorem 2.2]{Alpago-F-18}. An analogous  result is   Statement 1. (iii) of Theorem 3.1 in \cite{Ferrante-P-17} although referring to the specific case of all-pass functions.

\begin{lem}
{\em
Any solution $P$ of the Riccati equation \eqref{ch-AREP}
 corresponding to a $\Gamma\tp$-invariant subspace ${\mathcal S}$ can be expresses by the formula
\beq\label{ch-formadiP}
P=\left[(I-\Pi_{\mathcal S}) P_{+}^{-1} (I-\Pi_{\mathcal S})\right]\pinv
\eeq
where $\pinv$ denotes Moore-Penrose pseudoinverse and $\Pi_{\mathcal S}$ is the orthogonal projector
 onto the subspace ${\mathcal S}=\ker{P}$.
 }
  \end{lem}
  We are now ready to present our main result.
\begin{thm}
{\em 
 Let $P$ be an arbitrary solution of   \eqref{ch-AREP}. Then the  unique solution   $P_{\small Q}$  of  \eqref{ch-AREPq} such that $\ker(P)=\ker(P_{\small Q})$ can be expressed by the formula
 \beq \label{PQ}
{P}_{\small Q} =  [PP\pinv Q PP\pinv + P\pinv]\pinv 
\eeq
}
wich generalizes \eqref{Nonsing}.
 \end{thm}
 \proof  
 Since $(I-\Pi_{\mathcal S})$ projects onto the range space of  $P$, a basic property of the Moore-Penrose pseudoinverse \cite[P. 421]{Horn-J-85}  implies that $ (I-\Pi_{\mathcal S})=PP\pinv $ so that  (\ref{ch-formadiP})   can be rewritten
$P=\left[PP\pinv P_{+}^{-1} PP\pinv\right]\pinv$ and hence
\beq\label{ch-exprforppinv}
P\pinv= PP\pinv
 P_{+}^{-1} PP\pinv \,.
 \eeq
Now, since $P$ and $P_{\small Q}$ have the same kernel they also have the same image so that the orthogonal  projectors on this image may be written in two ways as:
\beq\label{ch-twoformsproj}
I- \Pi_{\mathcal S}=PP\pinv = P_{\small Q}P_{\small Q}\pinv.
\eeq
Thus, the analog of formula (\ref{ch-formadiP}) for $P_{\small Q}$ yields
\beq\label{ch-formadiPq}
P_{\small Q}=\left[(I-\Pi_{\mathcal S}) P_{\small Q +}^{-1} (I-\Pi_{\mathcal S})\right]\pinv=\left[PP\pinv P_{\small Q +}^{-1} PP\pinv\right]\pinv
\eeq
where $P_{\small Q +}$  is the only non-singular solution of (\ref{ch-AREPq}) (such a solution corresponds to the $\Gamma\tp$-invariant subspace $\{0\}$).\\
Hence, after inserting \eqref{Nonsing}, we get 
${P}_{\small Q}= [PP\pinv (Q + P_{+}^{-1})PP\pinv]\pinv, $
and, finally,  by using (\ref{ch-exprforppinv}) we obtain the following explicit expression for $P_{\small Q}$
depending only on $P$ and $Q$:
\beq
{P}_{\small Q}= [PP\pinv QPP\pinv + P\pinv]\pinv.
\eeq
\qed

Finally, let us consider two arbitrary $\Gamma\tp$ and $A$- invariant subspaces ${\script X}$ and ${\script Y}$ which  is to say two arbitrary zero and pole flipping transformations of the singularities of $W_0(z)$ or, equivalently, an arbitrary state space and causality configuration for the model \eqref{star}. Suppose we want to compute   the corresponding minimal spectral factor $W(z)$, or equivalently the corresponding model \eqref{star}. Let $P$ and $Q$ be the  solutions  of the Riccati equations \eqref{ch-AREP} and \eqref{ch-AREQ} corresponding to the invariant subspaces   ${\script X}$ and ${\script Y}$  and consider the left lower path in the commutative diagram \eqref{CD} so that the zero flipping is done after a pole flipping defined by $Q$.  The relevant Riccati solution ${P}_{\small Q}$ is given in formula \eqref{PQ} so that the desired realization of $W(z)$ 
can be explicitly written in closed form as
\beq \label{WQP-finale}
W_{\small Q}(z):= C_{\small Q} (zI-A_{\small Q})^{-1}B_{\small P_Q}+ D_{\small P_Q}\,
\eeq
where  
\beq \label{PBDFinalissimo}
\begin{aligned}
&
B_{\small P_Q}:=(B_QD_Q\tp +A_QP_QC_Q\tp ) (D_QD_Q\tp +C_QP_QC_Q\tp)^{-1/2}\,\\
& D_{\small P_Q}:= 
(D_QD_Q\tp+C_QP_QC_Q\tp)^{1/2}\,,
\end{aligned}
\eeq
$P_Q$ is given by (\ref{PQ}) and $A_Q,B_Q,C_Q,D_Q$ are given by (\ref{CDQ}).

Naturally, an analogous procedure would work by following the upper right  path; i.e.  computing first $P$ and then performing the appropriate pole flipping defined by $Q_{\small P}$.

\section*{Conclusion}
We have discussed the classification of general (not necessarily analytic) square spectral factors in terms of the solutions of  two  algebraic Riccati equations. We have also  described the construction of general spectral factors which  involve both zero- and  pole-flipping on an arbitrary reference spectral factor.

\end{document}